# CLUSTER-BASED LOAD BALANCING ALGORITHMS FOR GRIDS

Resat Umit Payli[1], Kayhan Erciyes[2], Orhan Dagdeviren[2]

[1]Indiana University-Purdue University, Purdue School of Eng. and Tech.
Computational Fluid Dynamics Lab., Indianapolis, Indiana 46202, U.S.A.
`rpayli@iupui.edu`
[2]Computer Engineering Department, Izmir University, Izmir, Turkey
`{kayhan.erciyes,orhan.dagdeviren}@izmir.edu.tr`

*ABSTRACT*

*E-science applications may require huge amounts of data and high processing power where grid infrastructures are very suitable for meeting these requirements. The load distribution in a grid may vary leading to the bottlenecks and overloaded sites. We describe a hierarchical dynamic load balancing protocol for Grids. The Grid consists of clusters and each cluster is represented by a coordinator. Each coordinator first attempts to balance the load in its cluster and if this fails, communicates with the other coordinators to perform transfer or reception of load. This process is repeated periodically. We analyze the correctness, performance and scalability of the proposed protocol and show from the simulation results that our algorithm balances the load by decreasing the number of high loaded nodes in a grid environment.*

## 1. INTRODUCTION

Future e-science applications will require efficient processing of the data [1] where storages and processors may be distributed among the collaborating researchers. A computational Grid consists of heterogenous computational resources, possibly with different users, and provide them with remote access to these resources [2, 3, 4] and it is an ideal computing environment for e-science applications [5, 6, 7]. The Grid has attracted researchers as an alternative to supercomputers for high performance computing. One important advantage of Grid computing is the provision of resources to the users that are locally unavailable. Since there are multitude of resources in a Grid environment, convenient utilization of resources in a Grid provides improved overall system performance and decreased turn-around times for user jobs [8]. Users of the Grid submit jobs at random times. In such a system, some computers are heavily loaded while others have available processing capacity. The goal of a load balancing protocol is to transfer the load from heavily loaded machines to idle computers, hence balance the load at the computers and increase the overall system performance. Contemporary load balancing algorithms across multiple/distributed processor environments target the efficient utilization of a single resource and even for algorithms targeted towards multiple resource usage, achieving scalability may turn out difficult to overcome.

A major drawback in the search for load balancing algorithms across a Grid is the lack of scalability and the need to acquire system-wide knowledge by the nodes of such a system to perform load balancing decisions. Scalability is an important requirement for Grids like NASA`s Information Power Grid (IPG) [9]. Some algorithms have a central approach, yet others require acquisition of global system knowledge. Scheduling over a wide area network requires *transfer* and *location* policies. Transfer policies decide *when* to do the transfer[10] and this is typically based on some threshold value for the load. The location policy [11] decides where to send the load based on the system wide information. Location policies can be *sender initiated* [12, 13, 14] where heavily loaded nodes search for lightly loaded nodes, *receiver initiated* [15] in which case, lightly-loaded nodes search for senders or *symmetrical* where both senders and receivers search for partners [16]. Some agent based and game theoretic approaches were also proposed previously [17, 18, 19, 20]. Load balancing across a Grid usually involves sharing of data as in an MPI (Message Passing Interface) *scatter* operation as in [21], [22]. MPICH-G2, is a Grid-enabled implementation of MPI that allows a user to run MPI programs across multiple computers, at the same or different sites, using the same commands that would be used on a parallel computer [23].

In this study, we propose a dynamic and a distributed protocol based on our previous work [24] with major modifications and detailed test results to perform load balancing in Grids. The protocol uses the clusters of the Grid to perform local load balancing decision within the clusters and if this is not possible, load balancing is performed among the clusters under the control of clusterheads called the *coordinators*. We show that the protocol designed is scalable and has favorable message and time complexities.

The rest of the paper is organized as follows: In Section 2, the proposed protocol including the coordinator and the node algorithms is described with the analysis. In Section 3, the implementation of the protocol using an example is detailed and test results using Indiana University Grid environment are analyzed in Section 4 and Section 5 has the concluding remarks along with discussions.

## 2. THE PROTOCOL

We extend the load balancing protocol [24] for Grids to achieve a more balanced load distribution [25]. We use the same *daisy* architecture shown in Fig. 1, which is shown to be more scalable for group communication among other well-known architectures [26]. In this architecture, coordinators are the interface points for the nodes to the ring and perform load transfer decisions on behalf of the nodes in their clusters they represent. They check whether load can be balanced locally and if this is not possible, they search for potential receivers across the Grid same as in [24]. Additionally, by using the advantage of a daisy architecture, a token circulates the coordinator ring and can carry the global load information to all coordinators when it is needed. By using this information, coordinators can distribute the load in a more balanced way and also know the time to finish. These extensions are detailed in Section 2.1.

In this paper, we categorize the *Load* to LOW, MEDIUM and HIGH classes. The node is LOW when it can accept load from other nodes. A node is HIGH loaded when it is detected to be higher than the *upper threshold* as defined in the previous protocol [24]. The main difference is in the MEDIUM load definition. A node is MEDIUM if it has a load above the maximum limit of LOW load and can accept load from other nodes to reach the *upper threshold* , which is called MEDIUM MAX. Thus, MEDIUM loaded nodes do not need to give loads to LOW loaded nodes since they are not overloaded. Beside the categorization of nodes, clusters can be classified as LOW, MEDIUM and HIGH as regards to the sum of all local node loads in order to get a global point of view. Based on these definitions, we state that the two main targets of the load balancing protocol is to distribute the excessive loads of HIGH loaded nodes and group of nodes which are called clusters, to LOW and MEDIUM nodes and clusters to reach a global stable load distribution and to consider the time and message complexities in this operation by eliminating unnecessary transmissions.

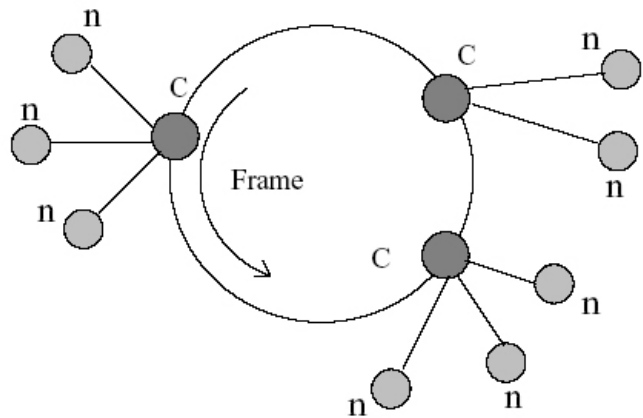

Figure 1. The Load Balancing Model for a Grid

## 2.1. Coordinator Algorithm

The primary functions of the coordinator is to monitor loads of the nodes in its cluster, initiate transfer of nodes from HIGH to LOW and MEDIUM nodes locally if possible and search for LOW and MEDIUM nodes across the Grid if there are no local matches. Its state diagram is shown in Fig. 2 and its pseudocode is given in Alg. 1.

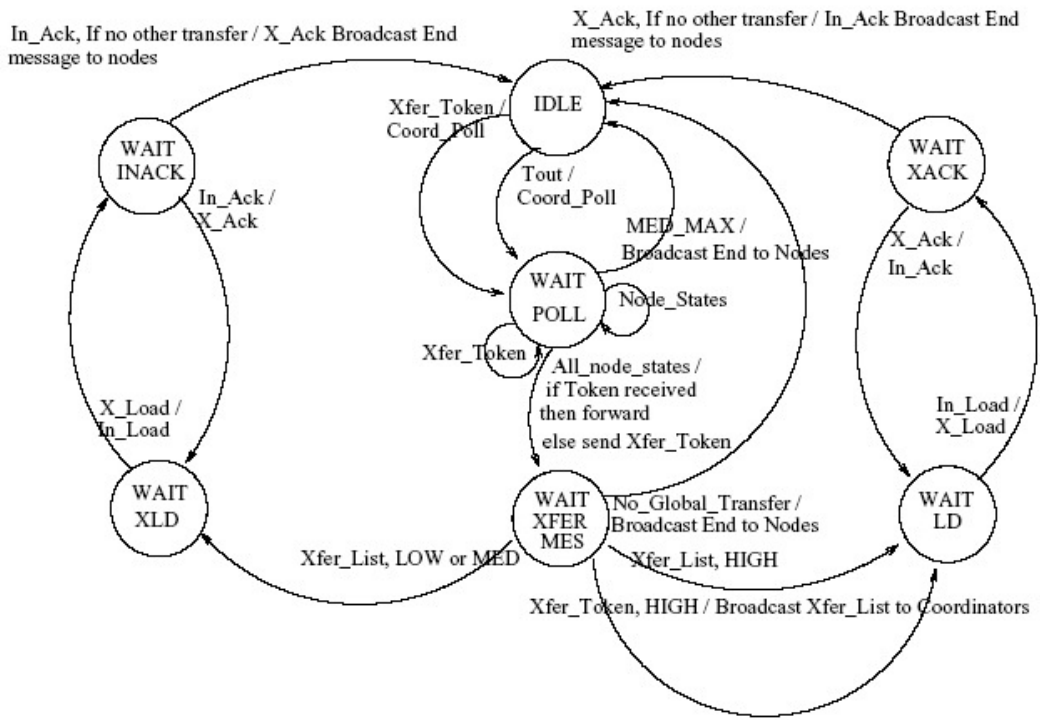

Figure 2. Coordinator Algorithm State Machine

**Algorithm 1** Load balancing algorithm for coordinator

1: **initially** $M$: set of all member nodes of the coordinator the node
2:     $C$: set of all coordinators
3:     $load_M$: received load array of coordinator from $M$
4:     $T_M$: total load of $load_M$
5:     $token\_received$: the token received state, initially equals to false.
6:     $next\_coordinator$: next coordinator (cluster leader) on the ring.
7:     **Legend** : □ State $\wedge$ $input\_message$ $\longrightarrow$ actions
8: **loop**
9:    □ IDLE $\wedge$ $Tout$ $\longrightarrow$ **multicast** $Coord\_Poll$ to $M$
10:     $current\_state \leftarrow$ WAIT_POLL
11:   □ IDLE $\wedge$ $Xfer\_Token$ $\longrightarrow$ **multicast** $Coord\_Poll$ to $M$
12:     $current\_state \leftarrow$ WAIT_POLL
13:     $token\_received \leftarrow$ true
14:   □ WAIT_POLL $\wedge$ $Node\_State$ $\longrightarrow$ **if** all loads are received from $M$ **then**
15:     **call** local_load_balance_procedure($load_M$)
16:     **if** $T_M$ is MEDIUM_MAX **then** **multicast** $End$ to $M$
17:       $current\_state \leftarrow$ IDLE
18:     **else send** $Xfer\_Token$ to the $next\_coordinator$
19:       $current\_state \leftarrow$ WAIT_XFER_MES
20:     **end if**
21:     **if** $token\_received$ = true **then forward** $Xfer\_Token$ to the $next\_coordinator$
22:     **end if**
23:   □ WAIT_XFER_MES $\wedge$ $Xfer\_Token$ originated from this coordinator $\longrightarrow$
24:     **calculate** global transfers similar to local_load_balance_procedure
25:     **if** no global transfer **then multicast** $End$ to $M$
26:       $current\_state$ *gets* IDLE
27:     **else multicast** $Xfer\_List$ to $C$
28:       **if** $T_M$ is HIGH **then** $current\_state \leftarrow$ WAIT_LD
29:       **else** $T_M$ is HIGH **then** $current\_state \leftarrow$ WAIT_XLD
30:       **end if**
31:     **end if**
32:   □ WAIT_XFER_MES $\wedge$ $Xfer\_List$ $\longrightarrow$ **if** no global transfers **then multicast** $End$ to $M$
33:     **end if**
34:     **if** $T_M$ is HIGH **then** $current\_state \leftarrow$ WAIT_LD
35:     **else** $current\_state \leftarrow$ WAIT_XLD
36:     **end if**
37:   □ WAIT_XACK $\wedge$ $X\_ACK$ $\longrightarrow$ **send** $In\_ACK$
38:     **if** no other transfer exists **then** $current\_state \leftarrow$ IDLE
39:     **else** $current\_state \leftarrow$ WAIT_LD
40:     **end if**
41:   □ WAIT_XLD $\wedge$ $X\_Load$ $\longrightarrow$ **send** $In\_Load$
42:     $current\_state \leftarrow$ WAIT_INACK
43:   □ WAIT_XACK $\wedge$ $In\_ACK$ $\longrightarrow$ **send** $X\_ACK$
44:     **if** no other transfer exists **then** $current\_state \leftarrow$ IDLE
45:     **else** $current\_state \leftarrow$ WAIT_XLD
46:     **end if**
47: **end loop**

Algorithm 1. Load Balancing Algorithm for Coordinator

```
Algorithm 2 Local load balance procedure
 1: initially: load[i]: the load value of i^th node
 2: input: load array
 3: for each load[i] do
 4:   if load[i] > MEDIUM_MAX then
 5:     excessive_load ← load[i]-MEDIUM_MAX
 6:     for each load[j] except j=i do
 7:       if excessive_load = 0 then
 8:         break the loop
 9:       end if
10:       if load[j] < MEDIUM_MAX then
11:         if MEDIUM_MAX - load[j] >= excessive_load then
12:           excessive_load ← 0
13:           transferable_load ← excessive_load
14:         else if
              then
15:           excessive_load ← excessive_load-(MEDIUM_MAX-load[j])
16:           transferable_load ← MEDIUM_MAX-load[j]
17:         end if
18:       end if
19:       load[i] ← load[i]-transferable_load
20:       load[j] ← load[j]-transferable_load
21:     end for
22:   end if
23:   create a new transfer T(from:i, to:j, load:transferable_load)
24:   start the transfer T
25: end for
26: output: load array
```

Algorithm 2. Local Load Balance Procedure

The coordinator of a cluster is activated periodically and starts the protocol by sending a *Coord Poll* message to every node in its cluster. It waits to receive all *Node State* messages from all ordinary nodes belonging to the same cluster. After receiving all nodes from its cluster, it checks whether there are any local matching nodes by executing the pseudocode in Alg. 2. If there is a match, it sends *Coord Xfer* message to HIGH node to initiate the local transfer. Before starting a local transfer, the coordinator node sets the load of both parties to end configuration. Also it calculates the overall cluster load by adding the loads of each node. If the cluster state is HIGH loaded, it must send or forward a *Xfer Token* message to its next coordinator on the ring. If it hasn't received a *Xfer Token* message yet, it sends a new *Xfer Token* message to its next coordinator otherwise picks up the received *Xfer Token* with the smallest *node id* and forwards it. *Xfer Token* message includes the overall cluster load and it is used for global load balancing.

After the *Xfer Token* of any coordinator is circulated across the ring, the originator node broadcasts a *Xfer List* message which consists of all overall cluster loads. It is important to note that the local transfers can occur in parallel with this event, since they would not affect the overall cluster loads. After receiving the *Xfer List*, all coordinators create the global transfers by an algorithm as in Alg. 2. If the coordinator has no global transfer or completed its global transfers and has finished all local transfers, it broadcasts an *End* message to its cluster in order to finish the load balancing operation.

## 2.2. Node Algorithm

The node algorithm is same as our previous protocol [24] with minor modifications. The node process replies by sending its load status to the coordinator in return for the *Coord Poll* message from the coordinator. If its load is HIGH, it waits for initiation of transfer from the coordinator.

When it receives the (*Coord Xfer*) message meaning transfer can be started, it sends the excessive load to the receiver sent by the coordinator and waits for an acknowledgement to confirm that the transfer was successful. This process is shown in Fig. 3 [24] and its pseudocode is given in Alg. 3. Different than the previous protocol, if the load of the node is LOW or MEDIUM, it will wait for a transfer from the HIGH node or wait the *End* message.

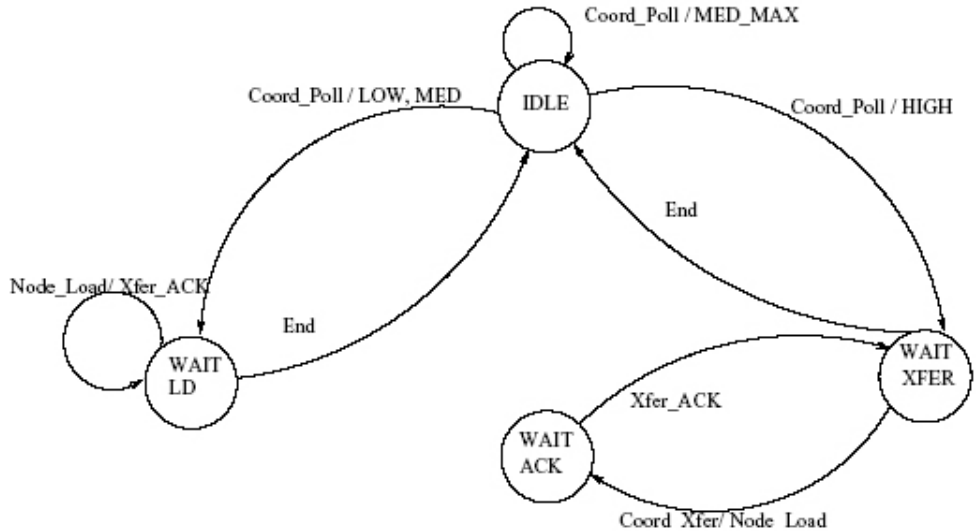

Figure 3. Node Algorithm State Machine

```
Algorithm 3 Load balancing algorithm for node
1: initially current_state: the current state of the node
2:     current_state ← IDLE
3:     load: node's load
4:     c: node' coordinator
5:     Legend : □ State ∧ input_message ⟶ actions
6: loop
7:     □ IDLE ∧ Coord_Poll ⟶ send load to c
8:        if load is HIGH then current_state ← WAIT_XFER
9:        else if load is MEDIUM_MAX then current_state ← IDLE
10:       end if
11:    □ WAIT_LD ∧ Node_Load ⟶ send Xfer_ACK
12:    □ WAIT_LD ∧ End ⟶ current_state ← IDLE
13:    □ WAIT_XFER ∧ Coord_Xfer ⟶ send Node_Load
14:       current_state ← WAIT_ACK
15:    □ WAIT_XFER ∧ End ⟶ current_state ← IDLE
16:    □ WAIT_ACK ∧ Xfer_ACK ⟶ current_state ← WAIT_XFER
17: end loop
```

Algorithm 3. Load Balancing Algorithm for Node

## 3. AN EXAMPLE OPERATION

In this section, an example operation of the protocol is described in a Grid network with 24 nodes and 4 clusters as shown in Fig. 4. Nodes 0, 6, 12 and 18 are the coordinators, other connected nodes are ordinary cluster members. The loads are represented with integer numbers from 1 to 15. LOW, MEDIUM and HIGH loads are divided into 1-5, 6-10, 11-15 intervals

258

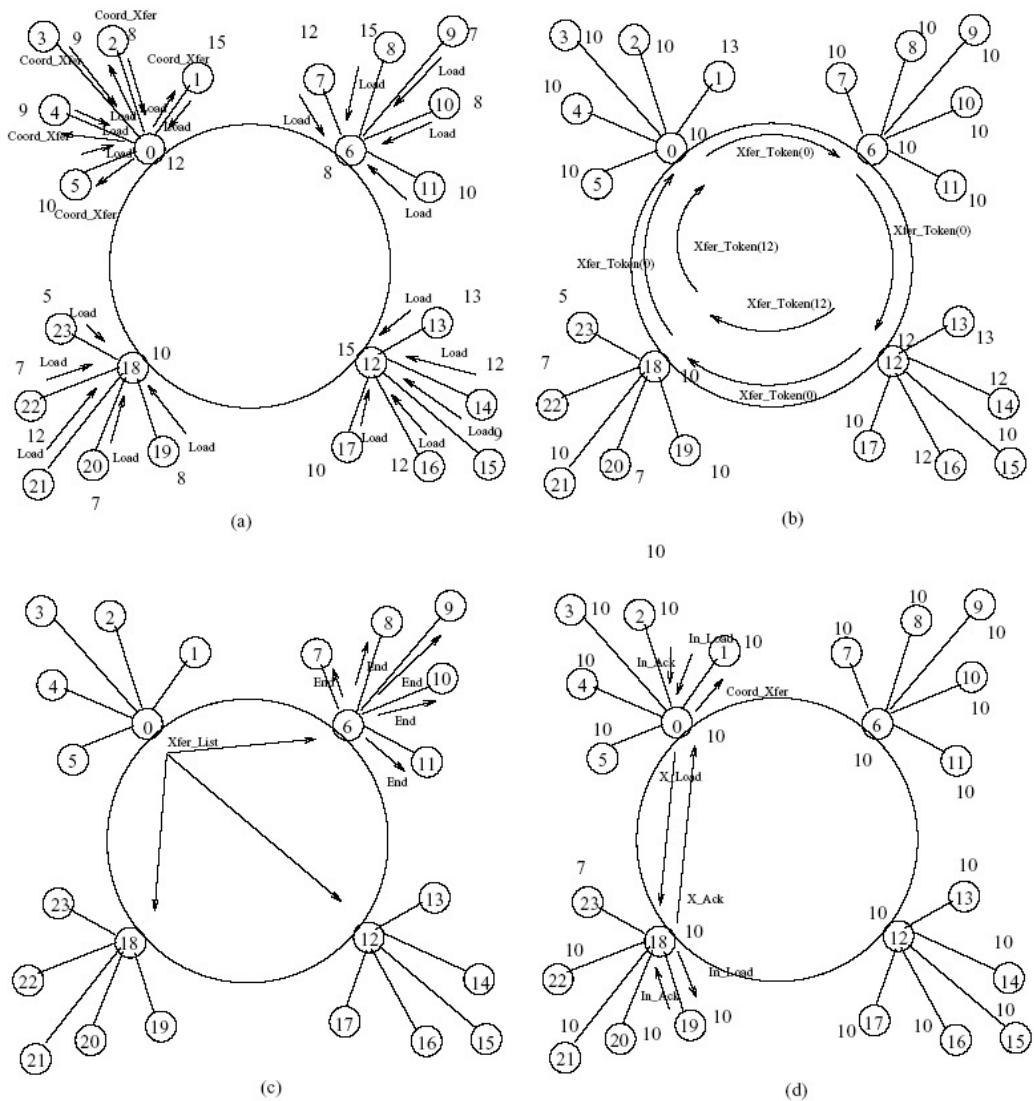

Figure 4. An Example Operation of the Protocol

respectively. The load of a cluster is the sum of all local nodes. Similar to the node loads, cluster loads are categorized as LOW, MEDIUM and HIGH. Each load of the node is depicted on top of its symbol in Fig. 4. The following are the sequence of events:

1. A timeout occurs in coordinator 0. It broadcasts *Coord Poll* messages to its nodes. Each node sends its load. After collecting all of the node states, coordinator 0 begins to make local load balancing. It sends *Coord Xfer* message to node 2 which is HIGH(15). Also coordinator 0 sends a *Coord Xfer* message to its own local node process. At the same time, coordinator 6, coordinator 12 and coordinator 18 wakes up and starts local load balancing same as coordinator 0, by executing pseudocode in Alg. 2. These operations are summarized in Fig. 4(a).

2. After local load balancing, the load values are depicted on top of symbol of each node in Fig. 4(b). All coordinators sum their local node's loads to make a decision about their cluster load state. Coordinator 0 and coordinator 12 find that their cluster load is HIGH. So they send *Xfer*

259

*Token* messages to their next coordinators on the ring which includes their cluster loads. Coordinator 12's *Xfer Token* message is dropped by coordinator 0, but coordinator 0's message circulates across the ring. While this token is traversing, each coordinator appends its total cluster load. These operations are summarized and depicted in Fig. 4(b).

3. After token is traversed, Coordinator 0 knows all the global information and broadcasts a *Xfer List* message to other nodes. After receiving the *Xfer List* message, coordinator 6, which's total load is equal to *MEDIUM MAX*, has no transfer so it broadcasts an *End* message to its nodes and return to *IDLE* state. Coordinator 0 and coordinator 12 which has excessive loads, calculates the local transfers by executing a pseudocode similar to local load balancing code in Alg. in 2. All of these message transmissions are shown in Fig. 4(c).

4. Coordinator 0 sends a *Coord Xfer* message to node 1 to start a load transfer. After node 1 receives the message, it sends the load to the coordinator 0, which will then be passed to coordinator 18. Coordinator 18 sends the load to the node 20. Node 20 sends an ack to coordinator 18 which will then be passed to coordinator 0 and node 1 sequentially. This operation is shown in Fig. 4(d). The other global transfers are (12 to 18, load:3), (12 to 18, load:3), (12 to 18, load:2). These transfers occur similar with this example and the final loads are gained. The final loads are shown in Fig. 4(d), on top of each node symbol. Lastly, coordinator 0, coordinator 12, coordinator 18 broadcasts *End* messages to their local cluster nodes.

## 3.1 Analysis

Let us assume that messages in the algorithms given in Alg. 1, Alg. 2 and Alg. 3 will arrive in finite time from a source to destination or multiple destinations. We also assume $k$, $d$ are upperbounds on the number of clusters and diameter of a cluster respectively.

Observation 1. *In coordinator algorithm given in Alg. 1 and node algorithm given in Alg. 3, each coordinator and node starts in IDLE state and should end in IDLE state.*

Lemma 1. *Coordinator algorithm given in Alg. 1 is free from deadlock and starvation.*

*Proof.* We prove the lemma by contradiction. From Observation 1, a coordinator starts executing the Alg. 1 in *IDLE* state and should end in *IDLE*. If there exists a deadlock or starvation, a coordinator should be any state other than *IDLE* after executing the algorithm. We assume the contrary that a coordinator is not in *IDLE* state at the end of the algorithm. In this case, a coordinator may be in any of *WAIT POLL*, *WAIT XFER MES*, *WAIT ACK*, *WAIT LD*, *WAIT LD*, *WAIT INACK*, and *WAIT XLD* states. We investigate all these cases below:

- *WAIT POLL*: A coordinator makes a state transition from *WAIT POLL* to *IDLE* or *WAIT XFER MES* state when it receives all *Node State*s from cluster members. From Observation 1, since all ordinary cluster member nodes start executing the algorithm in *IDLE* state, they reply the *Coord Poll* message with *Node State* as in FSM given in Fig. 3 and line 7 in Alg. 3. Thus a coordinator can not wait endlessly in this state.

- *WAIT XFER MES*: In this state, a coordinator node makes a state transition to *WAIT LD*, *WAIT XLD*, or *IDLE* state if it receives a *Xfer List* or its own circulated *Xfer Token* message. If it receives its circulated *Xfer Token* message, the coordinator multicasts a *Xfer List* message to all coordinators. Since at least one of the *Xfer Token* messages will be circulated the ring, a coordinator will receive its own *Xfer Token* or multicasted *Xfer List*. Thus a coordinator will make a state transition from this state.

260

International Journal of Computer Networks & Communications (IJCNC) Vol.3, No.5, Sep 2011

- *WAIT LD*: When a coordinator is in *WAIT LD* state, it waits for *In Load* message from ordinary nodes. To transfer the load, a coordinator sends *Coord Xfer* message to the destination node with *HIGH* load. When an ordinary node is *HIGH* loaded, it will make a state transition to *WAIT XFER* state and and it replies the *Coord Xfer* message with a *Node Load* message as shown in Fig. 3. So a coordinator will make a state transition from *WAIT LD* state to *WAIT XACK* since it receives the *In Load* message.

- *WAIT ACK*: A coordinator which is in this state will make a state transition to *IDLE* state upon receiving the last *X ACK* message. A *LOW* or *MEDIUM* loaded ordinary node will make a state transition to *WAIT LD* state when it receives *Coord Poll* message and it replies the *Node Load* message with a *Xfer ACK* message. So a coordinator will make a state transition to *IDLE* state in finite time.

- *WAIT XLD*: This case is similar to the case of a coordinator in *WAIT LD* state. When the coordinator in *WAIT XLD* state receives the *X Load* message it will make a state transition to *WAIT INACK* state.

- *WAIT INACK*: When the coordinator receives the last *In ACK* message, it makes a transition to IDLE state. This case is similar to the case of a coordinator in *WAIT ACK* state.

As explained from all the cases listed above, a coordinator node can not finish the algorithm in one of listed states. Thus we contradict with our assumption. The coordinator node will be *IDLE* state after executing the algorithm and coordinator algorithm is free from deadlock and starvation.

Lemma 2. *A coordinator will multicast End message to its cluster members before ending the execution of the algorithm.*

*Proof.* A coordinator starts the algorithm in *IDLE* state and makes a state transition to *WAIT POLL* state when a *Tout* occurs. From Lemma 1, all coordinators will be in *IDLE* state after the algorithm execution. As shown in Fig. 2 there are three transitions to the *IDLE* state and in each transition a coordinator multicasts *End* message.

Lemma 3. *Node algorithm given in Alg. 1 is free from deadlock and starvation.*

*Proof.* Assume the contrary. From Observation 1, a node should be in *IDLE* state at the end of the algorithm. Thus we assume that an ordinary node can be in *WAIT LD*, *WAIT ACK* or *WAIT XFER* state after the execution of the algorithm. When a node is LOW or MEDIUM loaded it makes a state transition from *IDLE* to *WAIT LD* state upon receiving *Coord Poll* as shown in Fig. 3. A node in this state will eventually receive *End* message from its coordinator as proved in Lemma 2 and will make a state transition to *IDLE* state. When a node is HIGH loaded it will make a state transition to *WAIT XFER* state and transfers its load upon receiving *Node Load* message. After it transfers the load, it will receive *Xfer ACK* message from its coordinator and makes a state transition to *WAIT XFER* state. When the coordinator sends a *End* message, this cyclic state transitions will finish and node will be in *IDLE* state at the end. We contradict with our assumption, node algorithm is free from deadlock and starvation.

Theorem 1. *Our hierarchial dynamic load balancing protocol is free from deadlock and starvation.*

*Proof.* Our protocol consists of coordinator and node algorithm. From Lemma 1 and Lemma 3 we proved that both algorithms are free from deadlock and starvation, thus theorem holds true.

Theorem 2. *The total number of messages for acquiring the global load knowledge at the worst case is k(1+(k-1)/2).*

*Proof.* A *Xfer Token* must circulate across the ring and must be transmitted back to originator node. *Xfer Token* with the smallest id will circulate, others will be dropped. So, at the worst case arrangement of coordinators, the sum of 1 to ($k$-1) *Xfer Token* messages, which is equal to $k(k-1)/2$, are transmitted. After originator node receives its token, it broadcasts $k$ *Xfer List* messages. Thus totally $k(1+(k-1)/2)$ messages are needed.

Corollary 1. *For the global load distribution algorithm, circulation of the token in daisy architecture eliminates at least half of messages with compared to broadcasting.*

*Proof.* When each coordinator broadcasts its load, totally ($k$-1)$k$ messages are needed. On the other side, from theorem 2, circulation of token requires $k(k-1)/2$ messages at the worst case.

Corollary 2. *The time for acquiring the global load knowledge is O(dk) where T is the average message transfer time between adjacent nodes.*

*Proof.* As shown in Theorem 2, at the worst case, $k(k-1)/2$ messages are needed for token circulation. However, since messages are transmitted in parallel, at the worst case $(2k-1)T$ time needed for token circulation. After originator node receives its token, it broadcasts $k$ *Xfer List* messages in $Td$ time. Totally $((2k-1)+d)T$ time is needed. Thus, the time for acquiring the global load knowledge is $O(k)$.

Theorem 3. *The time for a single load transfer is between 4dT+L and (4d+1)T+L where L is the actual average load transfer time.*

*Proof.* A node transfers its state to the coordinator in $d$ steps in parallel with the other nodes and assuming there is a match of LOW-HIGH nodes in the local cluster, the coordinator will send *X Load* message to the HIGH node in $d$ steps. Then there will be $L$ time for the actual load transfer. The HIGH and LOW(or MEDIUM) nodes also perform a final handshake to confirm delivery of load in $2d$ steps. The total minimum time for load transfer is then the sum of all of these steps which is $4dT+L$. In the case of a remote receiver, only 1 *Coord Xfer* message will be transmitted resulting in $(4d+1)T+L$ time.

Corollary 3. *The total number of messages exchanged for a single load transfer is O(d).*

*Proof.* As shown by Theorem 3, the maximum total number of messages required for a remote receiver will be $(4d+1)T+L$. Thus, the message complexity for the a single load transfer of the algorithm is $O(d)$.

## 4. PERFORMANCE EVALUATIONS

We implemented the load balancing algorithm with MPI library. We carried out our tests on Indiana University's BigRed Cluster. Indiana University's BigRed is a distributed shared-memory cluster, consisting of 768 IBM JS21 Blades, each with two dual-core PowerPC 970 MP

262

processors, 8GB of memory, and a PCI-X Myrinet 2000 adapter for high-bandwidth, low-latency MPI applications. In addition to local scratch disks, the BigRed compute nodes are connected via gigabit Ethernet to a 266TB GPFS file system, hosted on 16 IBM p505 Power5 systems. In our all experiments, each measurement is averaged with three identical measurement scenarios. We firstly measure the runtimes of the algorithm for varying nodes and cluster sizes. The node numbers are selected from 3 to 30. Since each node has 4 cores, the core numbers are varying from 12 to 120. The cluster sizes are selected as 3, 4 and 6. Fig. 5 shows that our algorithm is scalable when increasing core numbers from 12 to 120 and varying cluster sizes as 3,4 and 6. Since the nodes in the BigRed Cluster are identical, the cluster formation is simply achieved by considering the rank of the cores given by MPI. For example the cores 0, 1 and 2 are in the same the cluster in a system which is divided to the clusters having 3 cores each.

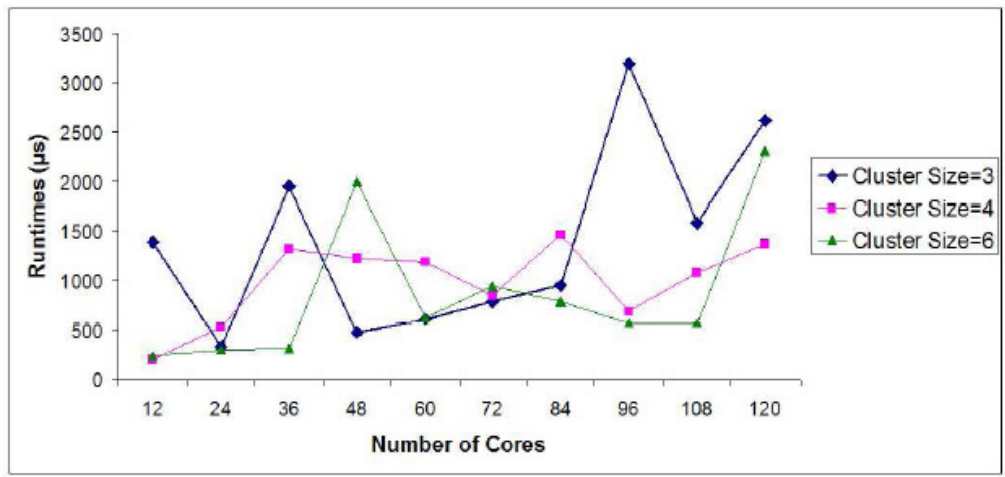

Figure 5. Runtimes of the Algorithm

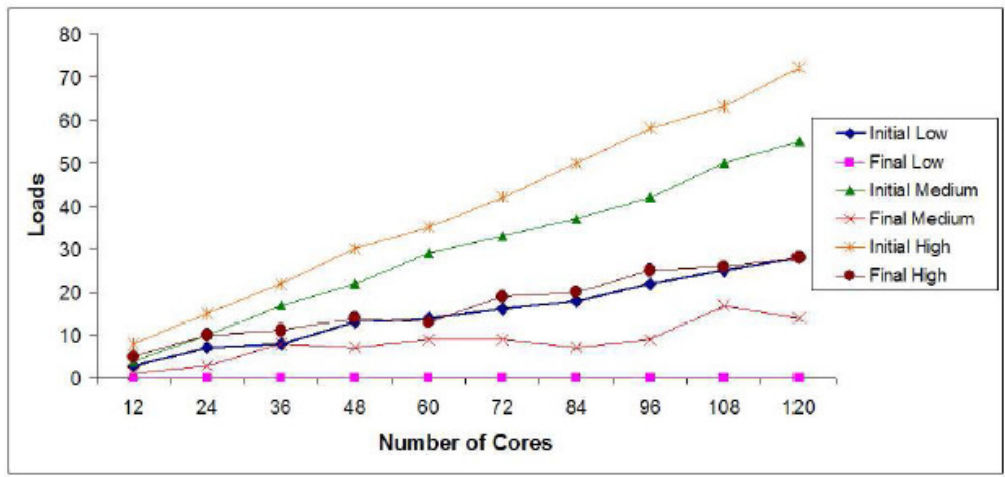

Figure 6. Number of High Loaded Cores w.r.t Total System Load

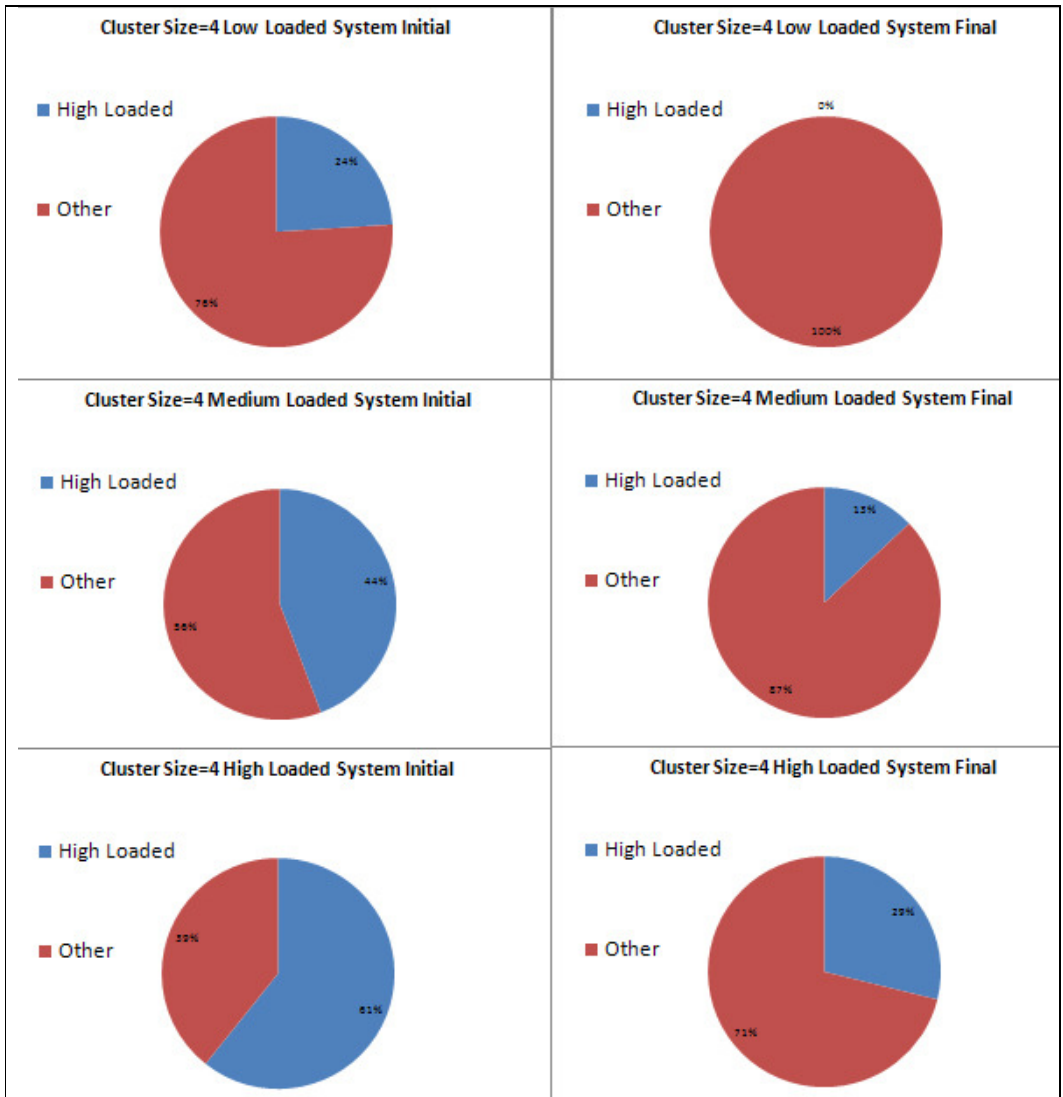

Figure 7. Percentage of High Loaded Cores w.r.t Total System Load

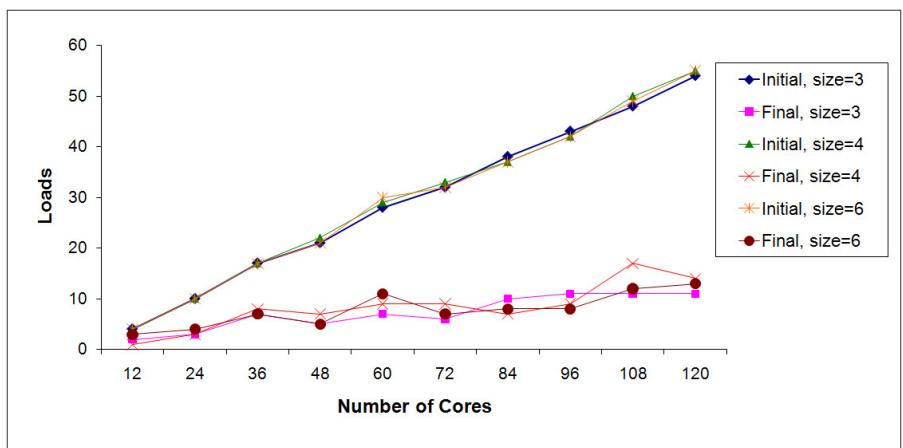

Figure 8. Number of High Loaded Cores w.r.t Cluster Size

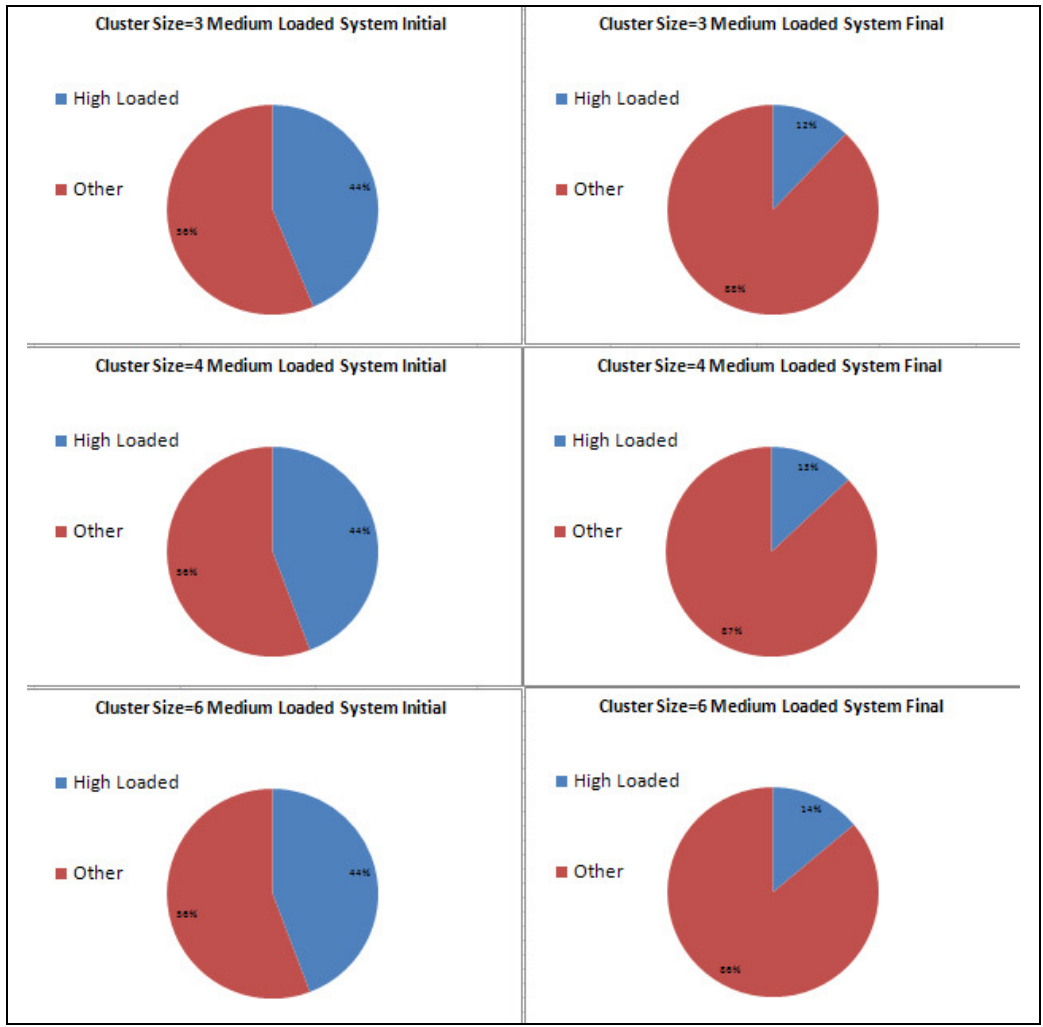

Figure 9.  Percentage of High Loaded Cores w.r.t Cluster Size

One important goal of the algorithm is to decrease the number of high loaded nodes in the system as much as possible. For the system with 4 cores in each cluster, we prepare low, medium and high loaded test scenarios randomly with different seeds. The loads are represented with the integer values. Between 5 and 20, each 5 integer interval belongs a load type low, medium and high respectively. Fig. 6 and Fig. 7 show the number and percentage of high loaded cores in initial state, before applying the algorithm, and in final state, after applying the algorithm. In the low loaded systems, all of the high loaded cores give their excessive loads to the other cores. In the medium loaded and high loaded systems more than half of the high loaded cores become medium loaded as shown in Fig. 6 and Fig. 7.

We also fix the system load as medium and change the cluster sizes to measure the number of high loaded cores with respect to various clustering schemes. Measurements in Fig. 8 and Fig. 9 shows that the curves in different clustering scenarios are similar which shows that our algorithm achieves both inter-cluster and intra-cluster balancing.

International Journal of Computer Networks & Communications (IJCNC) Vol.3, No.5, Sep 2011

Lastly, we measure the standard deviations of the loads before and after applying the algorithm in Fig. 10. In medium and especially high loaded systems, number of high loaded cores are much greater than the number of high loaded cores in low loaded systems. Generally, the standard deviations after applying the algorithm are smaller than half of the values before applying the algorithm in high and medium loaded systems as seen Fig. 10. In low loaded systems, since the number of transfers are smaller, the standard deviations after applying the algorithm are approximately 70% of the initial values. When we fix the system as medium loaded and vary the cluster size, the standard deviations after applying the algorithm are smaller than half of the initial values as seen in Fig. 11. This shows that the algorithm behaves stable under varying cluster sizes.

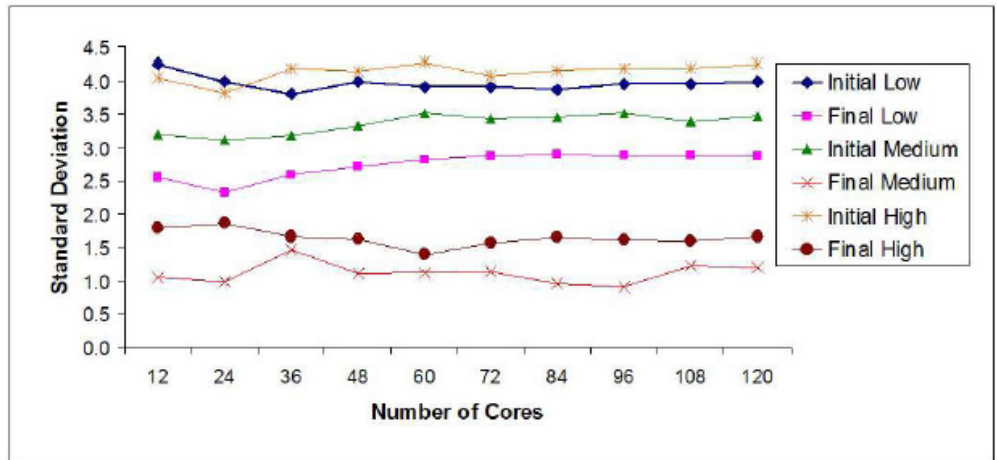

Figure 10. Standard Deviations w.r.t Total System Load

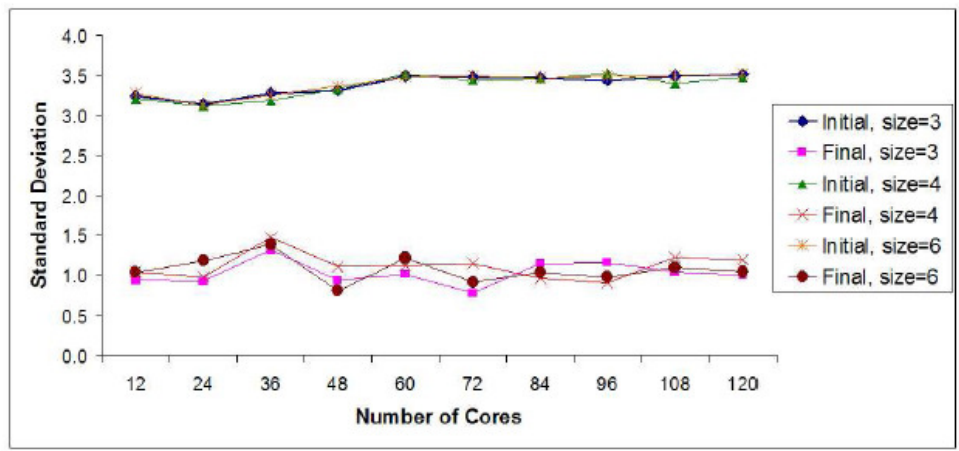

Figure 11. Standard Deviations w.r.t Cluster Size

## 5. CONCLUSIONS

We showed the design and implementation of a protocol for dynamic load balancing in a Grid. The Grid is partitioned into a number of clusters and each cluster has a coordinator to perform local load balancing decisions and also to communicate with other cluster coordinators across the Grid to provide inter-cluster load transfers, if needed. Our results confirm that the load balancing method is scalable and has low message and time complexities. We have not

266

addressed the problem of *how* the load should be transferred, and there are many research studies on this subject but we have tried to propose a protocol that is primarily concerned on *when* and *where* the load should be transferred. In fact, it may be possible not to transfer the load at all by employing copies of a subset of processes across the nodes in the Grid. The load transfer would require sending of some initial data only in this case. This is a future direction we are investigating.

The coordinators may fail and due to their important functionality in the protocol, new coordinators should be elected. Also, a mechanism to exclude faulty nodes from a cluster and add a recovering or a new node to a cluster are needed. These procedures can be implemented using algorithms as in [27] which is not discussed here. Our work is ongoing and we are looking into using the proposed model for real-time load balancing where scheduling of a process to a Grid node should be performed to meet its hard or soft deadline. The other area of concern discussed above would be the keeping the replicas of a subset of important and frequently used processes at Grid nodes to ease load transfer.

## ACKNOWLEDGEMENTS

This material is based upon work supported by the National Science Foundation under Grant No. ACI-0338618l, OCI-0451237, OCI-0535258, and OCI-0504075. This research was supported in part by the Indiana METACyt Initiative. The Indiana METACyt Initiative of Indiana University is supported in part by Lilly Endowment, Inc. This work was also supported in part by Shared University Research grants from IBM, Inc. to Indiana University.

**Authors**

**Resat Umit Payli**

Resat Umit Payli worked as a system administor in Ege University from 1978 to 1985. He was a research assistant in Ege University between from 1985 to 1987. He worked as a research assistant in Indiana University from 1987 to 1990. He was a research associate in Indiana University between from 1990 to 1992. He worked as a Software Engineer between from 1992 to 1999. He was a research associate in Indiana University from 1999 to 2012.

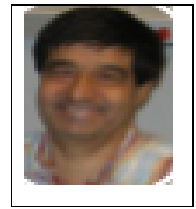

**Kayhan Erciyes**

Kayhan Erciyes received a BSc. degree in Electrical Eng. and Electronics from the University of Manchester, MSc. degree in Electronic Control Eng. from the University of Salford and a Ph.D. degree in Computer Engineering from Ege (Aegean) University. He was a visiting scholar at Edinburgh University Computer Science Dept. during his Ph.D. studies. Dr. Erciyes worked as visiting and tenure track faculty at Oregon State University, University of California Davis and California State University San Marcos, all in the U.S.A.

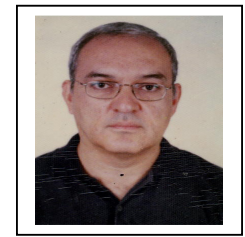

He also worked in the research and development departments of Alcatel Turkey, Alcatel Portugal and Alcatel SEL of Germany. His research interests are broadly in parallel and distributed systems and computer networks. More precisely, he works on distributed algorithms for synchronization in mobile ad hoc networks, wireless sensor networks and the Grid. Dr. Erciyes is the rector of the Izmir University in Izmir, Turkey.

**Orhan Dagdeviren**

Orhan Dagdeviren received the BSc. degree in Computer Eng. and MSc. degree in Computer Eng. from Izmir Institute of Technology. He received Ph.D. degree from Ege University, International Computing Institute. He is an assistant professor in Izmir University. His interests lie in the computer networking and distributed systems areas. His recent focus is on graph theoric middleware protocol design for wireless sensor networks, mobile ad hoc networks and grid computing.

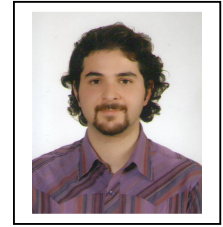